\documentclass[12pt]{article} 
\usepackage{a4wide}
\usepackage{latexsym}
\usepackage{epsfig}
\def\bq{\begin{eqnarray}}
\def\eq{\end{eqnarray}}

\def\eps{\varepsilon}

\begin{document}

\thispagestyle{empty}

\begin{flushright}
  NIKHEF/2001-013\\
  UPRF-2001-20\\
  September 2001\\
\end{flushright}

\vspace{18mm}
\begin{center}
  {\Large\bf Comparison of Phase Space Slicing and\\[0.3cm]
  Dipole Subtraction Methods for $\gamma^* \rightarrow Q\bar{Q}$}\\[1cm]
  {\sc Tim Oliver Eynck$^a$, Eric Laenen$^a$, Lukas Phaf$^{a}$, Stefan Weinzierl$^b$}\\
  \vspace{12mm}
  $^a${\it NIKHEF Theory Group,\\
       Kruislaan 409, 1098 SJ Amsterdam, The Netherlands,} \\
  \vspace{2.5mm}
  $^b${\it Dipartimento di Fisica, Universit\`{a} di Parma, \\
       INFN Gruppo Collegato di Parma, 43100 Parma, Italy} \\
\end{center}
\vspace{1cm}

\begin{abstract}
We compare the phase space slicing and dipole subtraction
methods in the computation of the 
inclusive and differential 
next-to-leading order cross sections 
for heavy quark production in the simple
process $\gamma^* \rightarrow Q\bar{Q}$.
For the phase space slicing method we study the
effects of improvement terms that remove restrictions 
on the slicing parameter $s_{\rm min}$. 
For the dipole method our comparison is a first check
on some of its counterterms involving massive quarks, derived recently.
In our comparison we address issues such as numerical accuracy
and efficiency.
\end{abstract} 
\newpage \setcounter{page}{1}

\section{Introduction}
\label{sec:introduction}

Fully differential QCD cross sections are important observables 
for studies at high-energy colliders. By allowing detector-specific
acceptance cuts on phase space variables they eliminate the need for extrapolation 
into unmeasured, and often also poorly
calculable regions, and thereby improve theory-experiment
comparisons. Reliable theoretical predictions for such differential
cross sections require the inclusion of at least next-to-leading order 
(NLO) QCD corrections. NLO calculations combine virtual one-loop 
corrections with the real emission contributions from unresolved partons.
These two parts are usually computed separately and each is
infrared divergent, only their sum is infrared finite.
NLO Monte-Carlo programs incorporate both pieces and allow the
simultaneous computation of many differential cross sections
for the particular reaction considered. 

However, these programs require 
that infrared singularities be eliminated before any numerical 
integration can be done. There are essentially two types of methods to effect
this cancellation.  The phase space slicing (PSS) method
\cite{Fabricius:1981sx,Kramer:1989mc,Baer:1989jg,Harris:2001sx} is 
based on approximating the matrix elements and the
phase space integration measure in boundary regions of phase space 
so integration may be carried out analytically. The subtraction method 
\cite{Ellis:1981wv,Kunszt:1989km,Mangano:1992jk} is based on adding and 
subtracting counter terms designed to approximate the real emission
amplitudes in the phase space boundary regions on the one hand, 
and to be integrable with respect to the momentum of an
unresolved parton.

For massless partons both methods are well-developed and have 
been widely used. A quite general formulation 
of phase space slicing has been given in
Ref.~\cite{Giele:1992vf,Giele:1993dj}. 
It was extended to include massive quarks and identified hadrons
in Ref.~\cite{Keller:1998tf}.

There exist two general formulations of the subtraction method.
One is the residue approach \cite{Frixione:1996ms},
the other the dipole formalism \cite{Catani:1997vz}.
Both can handle massless partons and identified hadrons
in the final and/or initial state. The extension of the dipole method to handle
massive quarks, using dimensional regularization, 
has been given recently in Ref.~\cite{Phaf:2001gc}.
An extension to photon radiation off massive fermions,
using small masses for infrared regularization,
was developed by Dittmaier in \cite{Dittmaier:1999mb}.
There are also hybrid methods 
\cite{Campbell:1998nn} that combine elements of the
slicing and subtraction methods such that both the 
resolved and unresolved contributions are numerically small 
and can be reliably integrated.

With general formulations of the phase space slicing and
dipole methods for massless and massive quarks now available,
it is interesting to compare their efficiency and accuracy.
In this paper we do this for (differential) ``cross sections'' for
heavy quark production in the process $\gamma^* \rightarrow
Q\bar{Q}$. This case is of course very simple
but also generic for more complicated processes.
In the case of the NLO cross section for $t\bar{t}H$ production
\cite{Reina:2001sf,Beenakker:2001rj} it was
recently verified \cite{Beenakker:2001rj} that the
slicing method and a somewhat differently phrased
dipole method \cite{othermassdipole} agreed.

Our results using the dipole subtraction 
method represent the first numerical implementation of some of 
the subtraction terms computed in \cite{Phaf:2001gc}.
While the dipole method is exact,
the PSS method requires the introduction of a theoretical
resolution parameter $s_{\mathrm{min}}$, usually required
to be quite small. We include improvement terms in the
PSS method \cite{Kilgore:1997sq}, and study their effect
of removing restrictions on the 
size of the phase space slicing cutoff.
This is especially important for cross sections involving heavy 
quark production, and allows for a free choice of slicing
parameter without reference to the heavy quark mass,
a prerequisite for considering the high-energy
or zero-mass limit.

The paper is structured as follows. In section 2 we
compute the fully differential cross section
for $\gamma^* \rightarrow Q \bar{Q}$ using the PSS method, 
including the improvement terms.
In section 3 we compute this cross section with the
dipole subtraction method. In section 4 we present 
a numerical comparison of the two methods, followed by our conclusions.

\section{Phase Space Slicing}
\label{sec:phase-space-slicing}

We consider the process 
\begin{equation}
  \label{eq:46}
  \gamma^* (q) \rightarrow Q(p_1) + \bar{Q}(p_2)\,,
\end{equation}
with $p_1^2 = p_2^2 = m^2$. The NLO corrections 
involve virtual corrections to (\ref{eq:46}) and the gluon bremsstrahlung
reaction
\begin{equation}
  \label{eq:17}
    \gamma^* (q) \rightarrow Q(p_1) + \bar{Q}(p_2) + g(p_3)\,,
\end{equation}
with $p_3^2=0$. We define the invariants
\begin{equation}
\label{eq:1}
s_{ij} \equiv 2 p_i \cdot p_j\,, \qquad
\tilde{s}_{ij} \equiv \left(p_i + p_j\right)^2 \ .
\end{equation}
The final state phase space for the 3 parton contribution
is divided into ``hard'' and ``soft'' regions.
The hard region, in which all 3 final state particles in
(\ref{eq:46}) are resolved, is defined such that 
$s_{13} > s_{\rm min}$ or $s_{23} > s_{\rm min}$.
(In an appendix we discuss this definition when more
than one color structure is present).
The complementary region is soft. Let us 
review the approximations involved in PSS, following
\cite{Kilgore:1997sq}. The 3 parton contribution to the 
fully differential decay can be written schematically as
\begin{eqnarray}
\label{eq:2}
d\,\Gamma_3 &=& |{\cal M}_3|^2\times  d\,\mbox{PS}_3 \nonumber \\
            &=& \left(|{\cal M}_3|^2\times (1-\theta_s)
               +|{\cal M}_3|^2\times\theta_s)\right)
               \times  d\,\mbox{PS}_3 \nonumber \\
        &=& |{\cal M}_3|^2\times (1-\theta_s) d\,\mbox{PS}_3
               +\theta_s\times
               \left(T_1(\theta_s)+T_2(\theta_s)+T_3(\theta_s)\right)\ ,
\end{eqnarray}
where $|{\cal M}_3|^2$ is the exact matrix element squared, 
and $d\,\mbox{PS}_3$ denotes the exact 3 particle phase
space measure. Note we do not consider the effect of jet-algorithms
here (they are implicit in the definition of the phase space).
The slicing of phase space is indicated
by the symbol $\theta_s$, which is 0
in the hard phase space region and 1 in the
soft region. $T_1$ is given by
\begin{eqnarray}
\label{eq:T1}
T_1(\theta_s) &=& 
S\,|{\cal M}_2|^2\times  d\,\mbox{PS}_{\mathrm{soft}}\ d\,\mbox{PS}_2
\nonumber \\
    &=& R(\theta_s)\,|{\cal M}_2|^2\times  d\,\mbox{PS}_2\ ,
\end{eqnarray}
and represents the integral of the approximate matrix element $|{\cal
M}_3|^2 \rightarrow S\, |{\cal M}_2|^2$ over the approximate phase
space $d\,\mbox{PS}_3\rightarrow d\,\mbox{PS}_{\rm soft}\
d\,\mbox{PS}_2$.  The resolution factor $R(\theta_s)$ is
independent of the hard scattering and can be calculated analytically
for a wide range of multiparton processes \cite{Giele:1992vf,Giele:1993dj,Keller:1998tf}.  
$T_2$ is
given by
\begin{equation}
\label{eq:T2}
T_2(\theta_s) = \left( |{\cal M}_3|^2 - S\,|{\cal M}_2|^2\right)
\times  d\,\mbox{PS}_3\ ,
\end{equation}
and represents the integral over the exact 3-particle phase space phase space of the
difference between the true matrix element and the approximate matrix
element. $T_3$ is given by
\begin{equation}
\label{eq:T3}
T_3(\theta_s) = S\,|{\cal M}_2|^2\left(  d\,\mbox{PS}_3
      - d\,\mbox{PS}_2\,d\,\mbox{PS}_{\rm soft}\right)\ ,
\end{equation}
and represents the difference between the integrals of the approximate
matrix element over the true and approximate unresolved phase space.
Note that $T_1$ contains the soft
and collinear divergences needed to cancel the singularities of the
virtual term, while $T_2$ and $T_3$ are finite and vanish as 
the domain of support for $\theta_s$ is taken to zero.

\subsection{Matrix element}
\label{sec:matrix-element}

The matrix elements for the NLO cross section for process
(\ref{eq:46}) are not very complicated, so we can be explicit.  At
lowest order we have
\begin{eqnarray}
  \label{eq:47}
  d\Gamma_2 &=& \frac{1}{3} \frac{1}{2\sqrt{s}} 
 N e_q^2 e^2 \left(8 m^2 +4 s\right)  d\mathrm{PS}_2 \nonumber \\
 &=&  \frac{1}{3} \frac{1}{2\sqrt{s}} 
\left|{\cal M}_{\rm Born}\right|^2d \mathrm{PS}_2 
\end{eqnarray}
where $N$ is the number of colors, $e_q$ the fraction of the
elementary charge $e$ of the heavy quark, $m$ its mass, $s = q^2$ and
\begin{equation}
  \label{eq:48}
  d\mathrm{PS}_2 = \frac{1}{(2\pi)^2}
    \frac{d^3 p_1}{2 E_1} \frac{d^3 p_2}{2 E_2}
   \delta^{(4)}(q-p_1-p_2)\,.
\end{equation}
Note that in NLO approximation $|{\cal M}_2|^2$ in Eq.~(\ref{eq:T1})
is $\left|{\cal M}_{\rm Born}\right|^2$.  At ${\cal O}(\alpha_s)$
there are virtual and real emission contributions.  The PSS method
separates the latter into hard and soft contributions. The (spin and
color-summed) matrix element for the real emission process
(\ref{eq:17}) is
\begin{equation}
\label{eq:15}
\left|{\cal M}_3\right|^2 = 16 \, e_q^2 e^2 g_s^2 N C_F\, I_R
\end{equation}
with $g_s$ the strong coupling, $C_F = (N^2-1)/2N$ and
\begin{eqnarray}
\label{eq:16}
I_R & = & 
- {{m^2 s_{23}}\over{s_{13}^2}}
- {{m^2 s_{12}}\over{s_{13}^2}}
- {4{m^4}\over{s_{13}^2}}
+ {4{m^2 s_{12}}\over{s_{13} s_{23}}}
+ {{s_{12}^2}\over{s_{13} s_{23}}}
+ {{s_{23}}\over{2s_{13}}}
+ {{s_{12}}\over{s_{13}}} \nonumber \\ 
&& \ 
- {{m^2}\over{s_{13}}}
- {{m^2 s_{13}}\over{s_{23}^2}}
+ {{s_{13}}\over{2 s_{23}}}
- {{m^2 s_{12}}\over{s_{23}^2}}
- {4{m^4}\over{s_{23}^2}}
+ {{s_{12}}\over{s_{23}}}
- {{m^2}\over{s_{23}}}
\end{eqnarray}
In the $T_1$ term (\ref{eq:T1}) the eikonal approximation of the exact
matrix element is used. 
The integral over $d \mathrm{PS}_{\mathrm{soft}}$ is
then performed analytically and added to the virtual corrections.  The
approximated matrix element in the soft region (\ref{eq:T1}) is
\begin{equation}\label{eq:9}
S\,\left|{\cal M}_2\right|^2 = 16 \, e_q^2 e^2 g_s^2 N C_F \, I_S \ ,
\end{equation}
where
\begin{equation}
\label{eq:10}
I_S  =  
- {{m^2 s_{12}}\over{s_{13}^2}}
+ {4{m^2 s_{12}}\over{s_{13} s_{23}}}
- {{m^2 s_{12}}\over{s_{23}^2}}
- {4{m^4}\over{s_{13}^2}}
- {4{m^4}\over{s_{23}^2}}
+ {{s_{12}^2}\over{s_{13} s_{23}}} \ . 
\end{equation}
Note that the difference of (\ref{eq:16}) and (\ref{eq:10}) which
enters the $T_2$ term (\ref{eq:T2}), is finite in the limit
$s_{13},s_{23} \rightarrow 0 $.

The result of integrating (\ref{eq:9}) over $d
\mathrm{PS}_{\mathrm{soft}}$ is given in \cite{Keller:1998tf}, and
when added to the virtual contributions, gives the following finite
expression for the 2 particle ${\cal O}(\alpha_s)$ differential cross
section for process (\ref{eq:46})
\begin{equation}
  \label{eq:6}
    d\Gamma_2 = \frac{1}{3} \frac{1}{2\sqrt{s}} 
\left( \left|{\cal M}\right|_{\mathrm{soft}}^2
\,+\, \left|{\cal M}\right|_{\mathrm{virt}}^2 \right)
 d\mathrm{PS}_2\,,
\end{equation}
with
\begin{eqnarray}
\label{eq:12}
\left|{\cal M}\right|_{\mathrm{soft}}^2 &=& \frac{\alpha_s C_F}{\pi}\Bigg[
\frac{1}{\epsilon}\left(1+\left(1-{{2m^2}\over{s}}\right) {{\ln x}\over{\beta}}\right)
\Bigg]\,C_\epsilon\,\left|{\cal M}_{\rm Born}\right|^2 \nonumber \\
&\ &\qquad + \frac{\alpha_s C_F}{\pi}\Bigg[
-2 \left(1+\left(1-{{2m^2}\over{s}}\right) {{\ln x}\over{\beta}}\right) \left(\ln x - \ln \left({{s}\over{s_{\rm min}}}\right) - \ln \beta \right)\nonumber\\
&\ & -2\left( \ln \left(1-x\right)+\ln \left(1+x\right)-\ln x \right) \nonumber \\
&\ & +1
-{{\ln x}\over{\beta}} \left(1-{{2m^2}\over{s}}\right) \left(1+2 \ln {{\left(1-x\right)\left(1+x\right)}\over{x}}\right) \nonumber \\
&\ & +{{1}\over{2 \beta}} \left(1-{{2m^2}\over{s}}\right)\left({\rm Li}_2\left(1-{{1}\over{x^2}}\right) - {\rm Li}_2 \left(1-x^2\right)\right)
-\beta \nonumber \\
&\ & +{{m^2 }\over{s \beta}} \ln x \left({{1-x^2}\over{x}}+{{s}\over{m^2}}\left(1-{{2m^2}\over{s}}\right) \ln x \right)
+{{\ln^2 x}\over{2 \beta}}\left(1-{{2m^2}\over{s}}\right)\,.
\Bigg]\left|{\cal M}_{\rm Born}\right|^2 \nonumber \\
&\ &\qquad + \frac{\alpha_s C_F}{\pi} N e_q^2 e^2 \Bigg[
\left( 1+\left(1-{{2m^2}\over{s}}\right) {{\ln x}\over{\beta}}\right) \left(-4s\right)
\Bigg]
\end{eqnarray}
and
\begin{eqnarray}
\label{eq:13}
\left|{\cal M}\right|_{\mathrm{virt}}^2 &=& -\frac{\alpha_s C_F}{\pi}\Bigg[
\frac{1}{\epsilon}\left(1+\left(1-{{2m^2}\over{s}}\right) {{\ln x}\over{\beta}}\right)
\Bigg]\,C_\epsilon\,\left|{\cal M}_{\rm Born}\right|^2 \\
&\ &\qquad + \frac{\alpha_s C_F}{\pi}N e_q^2 e^2\Bigg[ -4s-16m^2 
- \frac{m^4}{s\beta}\left( 32 {\rm Li}_2\left(x\right) + 64  \zeta_2 \right)
 \nonumber \\
&&\ + \frac{s}{\beta}\left( 8 {\rm Li}_2\left(x\right) + 16  \zeta_2\right) 
+ \frac{1}{\beta}\ln^2\left(x\right) \left(8 \frac{m^4}{s}-2s\right) \nonumber \\
&& \ - \beta \ln x \left(6 s + 8 m^2\right)
+{{\ln x}\over{\beta}} \left(-32 \frac{m^4}{s} \ln \left(1-x\right) + 8 s \ln\left(1-x\right)
+4s -8m^2\right)\Bigg] \nonumber
\end{eqnarray}
Here $C_\epsilon = (4\pi\mu^2/m^2)^\epsilon/\Gamma(1-\epsilon)$,
$\beta = \sqrt{1-4m^2/s}$ and $ x = ({1-\beta})/(1+\beta)$. We
have written the divergent contributions explicitly, even though they
cancel between the soft and virtual contribution, so that the method
independent (virtual) and method dependent (soft) terms can be easily
read off. In particular we can obtain the results within the dipole
method by replacing the soft contribution with the integrated dipole terms.
Note the logarithmic dependence on the slicing parameter $s_{\rm min}$
in the finite soft contribution.

\subsection{Phase space}
\label{sec:phase-space}

The spin-summed squared matrix elements of the previous section are functions of the
final state momenta only via the invariants $s_{12},s_{13},s_{23}$. The
exact 3 particle phase space 
\begin{equation}
  \label{eq:8}
  d\mathrm{PS}_3 = \frac{1}{(2\pi)^5}
    \frac{d^3 p_1}{2 E_1} \frac{d^3 p_2}{2 E_2} \frac{d^3 p_3}{2 E_3}
   \delta^{(4)}(q-p_1-p_2-p_3)
\end{equation}
may be parametrized in terms of these invariants (after integrating
over all remaining variables)
\begin{eqnarray}
\label{eq:3}
  d\mathrm{PS}_3 =  \frac{1}{4s} \frac{1}{32 \pi^3} d s_{12}\,d s_{13}\,d s_{23}\,
  \delta(s - s_{12} -s_{13} -s_{23} -2m^2)\,.
 \end{eqnarray}
The integration limits of $s_{23}$ at fixed $s_{13}$ are
\begin{equation}
s_{23}^\pm = {{1}\over{2 \left(s_{13}+m^2\right)}} \left(-s_{13}\left(s_{13} - s + 2m^2\right) \pm s_{13} \sqrt{s_{13}^2 - 2 s_{13} s - 4 s m^2 + s^2}\right) \ .
\label{eq:4}
\end{equation}
\begin{figure}[!b]
\begin{center}
\epsfig{file=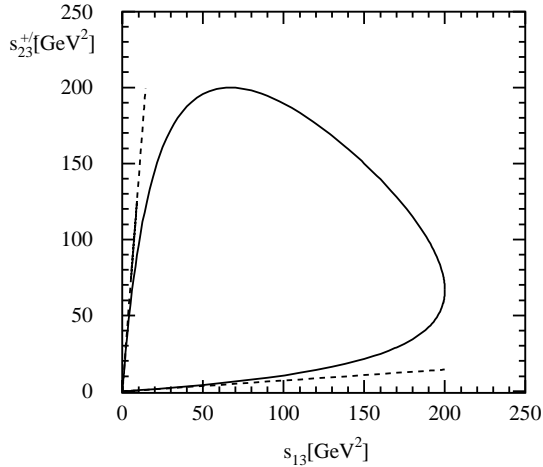,%
bbllx=50pt,bblly=120pt,bburx=285pt,bbury=460pt,angle=270,width=8.25cm}
\caption[dum]{\label{dahlitz-fig1}{\small{
Dahlitz plot for $s_{23}^\pm$ as a function of $s_{13}$ 
for exact (eq.(\ref{eq:4}), solid)
and eikonal (eq.(\ref{eq:11}), dashed) phase space boundaries 
at $m=5$ GeV and $s=400$ GeV$^2$.
 }}}
\end{center}
\vspace*{-0.5cm}
\end{figure}
The limits of $s_{13}$ at fixed $s_{23}$ are found by exchanging 
the indices $13$ and $23$. Setting $s_{23}^+ = s_{23}^-$ we find 
the maxima of these two invariants 
\begin{equation}
\label{eq:5}
s_{13}^{\rm max}= s_{23}^{\rm max} = s - 2 m \sqrt{s} \ .
\end{equation}
In the soft (eikonal) approximation, the limits for $s_{23}$ simplify to
\begin{eqnarray}
\label{eq:11}
s_{23}^{\pm,{\rm eik}} & = & {{1}\over{2 m^2}} \left(-s_{13}\left(2m^2 - s \right) \pm s_{13} \sqrt{s^2 - 4 s m^2 }\right) \nonumber \\
& = & s_{13}
\left( \frac{s-2m^2}{2m^2} \pm \frac{s}{2m^2}\beta\right) \ . 
\end{eqnarray}
The phase space boundaries for the exact and approximate cases 
are given by the Dahlitz plot in Fig.~\ref{dahlitz-fig1}.

\subsection{Results for PSS}
\label{sec:results-pss}
\begin{figure}[!h]
\begin{center}
\epsfig{file=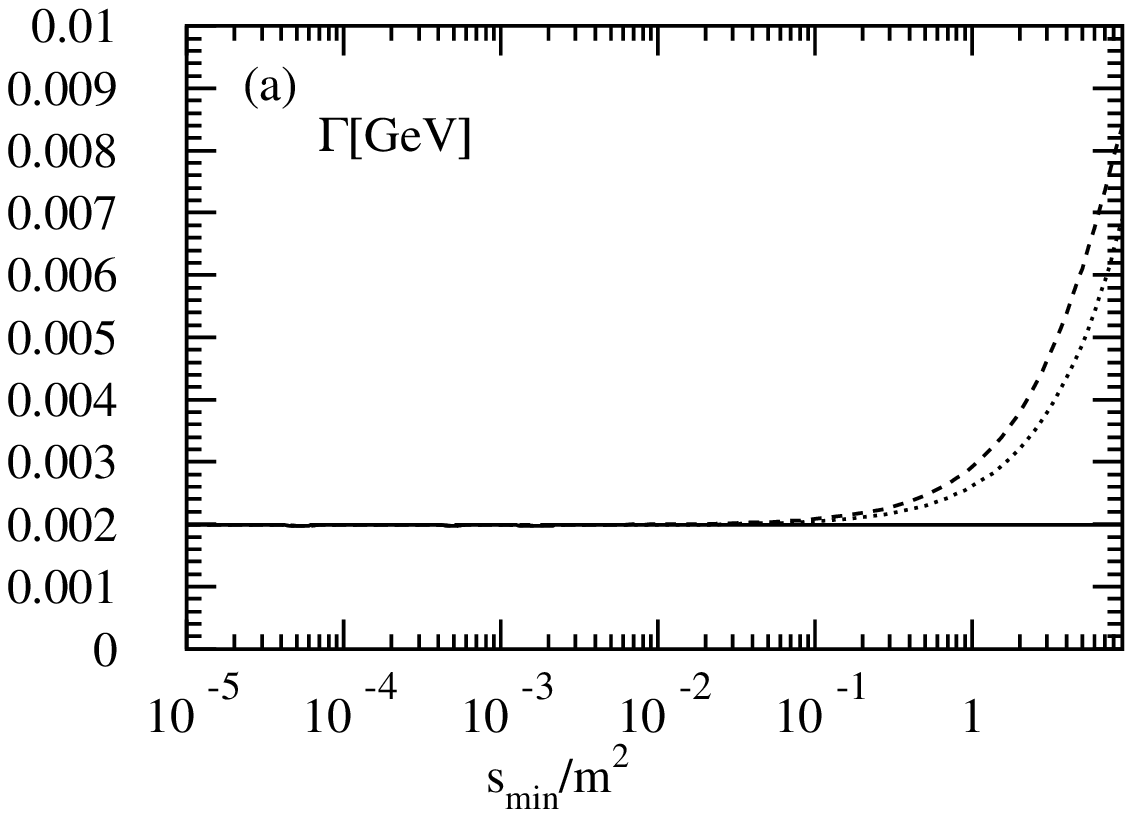,%
bbllx=50pt,bblly=120pt,bburx=285pt,bbury=450pt,angle=270,width=8.25cm}
\epsfig{file=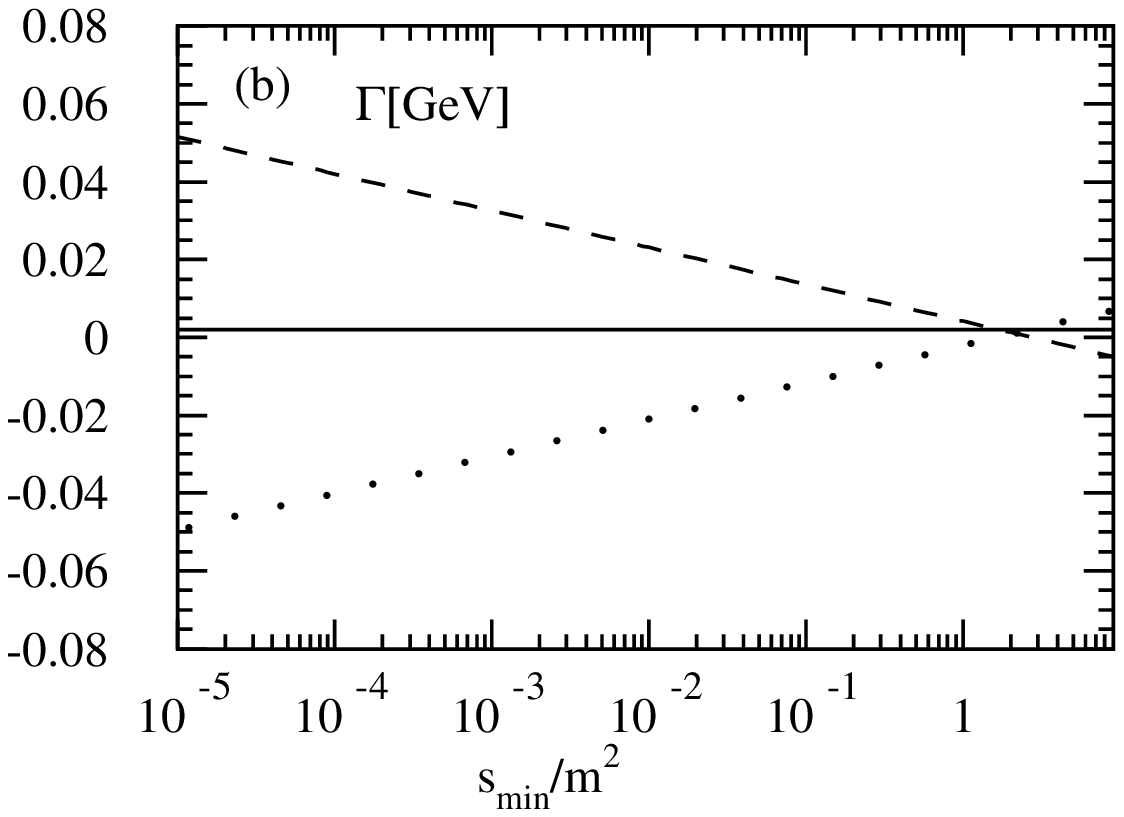,%
bbllx=50pt,bblly=120pt,bburx=285pt,bbury=450pt,angle=270,width=8.25cm}
\caption[dum]{\label{vary_smin-fig1}{\small{
      (a)
The $s_{\rm min}/m^2$  
dependence of the one-loop corrections to 
$\Gamma(s,m^2)$, when including
the $T_1$ (dotted), $T_1+T_2$ (dashed), and 
$T_1+T_2+T_3$ (solid) contributions.
      (b) 
The $s_{\rm min}/m^2$ 
dependence of the one-loop corrections to 
$\Gamma(s,m^2)$ for the soft$+$virtual (spaced dotted) and
the real emission (spaced dashed) final state contributions as well as 
their sum
(solid) in the $T_1+T_2+T_3$ approximation.
}}}
\end{center}
\end{figure}

We now show some results for the fully inclusive cross 
section, as well as some differential
distributions for process (\ref{eq:46}). 
We study what effect including the $T_i$ 
contributions has on the $s_{\rm min}$ dependence of the results, and
shall see that including all $T_i$ removes all $s_{\rm min}$ dependence.  
We use as default values $s=400$ GeV$^2$, and $m=5$ GeV.
Figure \ref{vary_smin-fig1}a shows that,
for the inclusive cross section, not
including all $T_i$ leads to $s_{\rm min}$ dependence (in fact
the $T_2$ worsens the $s_{\rm min}$ dependence slightly here), 
but including $T_2$ and $T_3$ relaxes
all constraints on this parameter.
This, however, comes at the expense of potentially lower 
numerical accuracy, particularly for the differential 
distributions to be considered below.
The inclusion of the $T_3$ term in particular requires 
a larger number of points in the Monte Carlo integration
than using $T_1$ alone, to achieve a given accuracy.
In practice, therefore, it is common to use only the $T_1$ in a PSS calculation,
with an $s_{\rm min}$ value small enough for the combined $T_2+T_3$ contribution to be negligible.
One must however be careful not to choose $s_{\rm min}$ so small that numerical 
inaccuracies result from the large opposite sign soft$+$virtual and 
real emission contributions, as illustrated by Figure \ref{vary_smin-fig1}b. 

\begin{figure}[hbtp]
\begin{center}
\epsfig{file=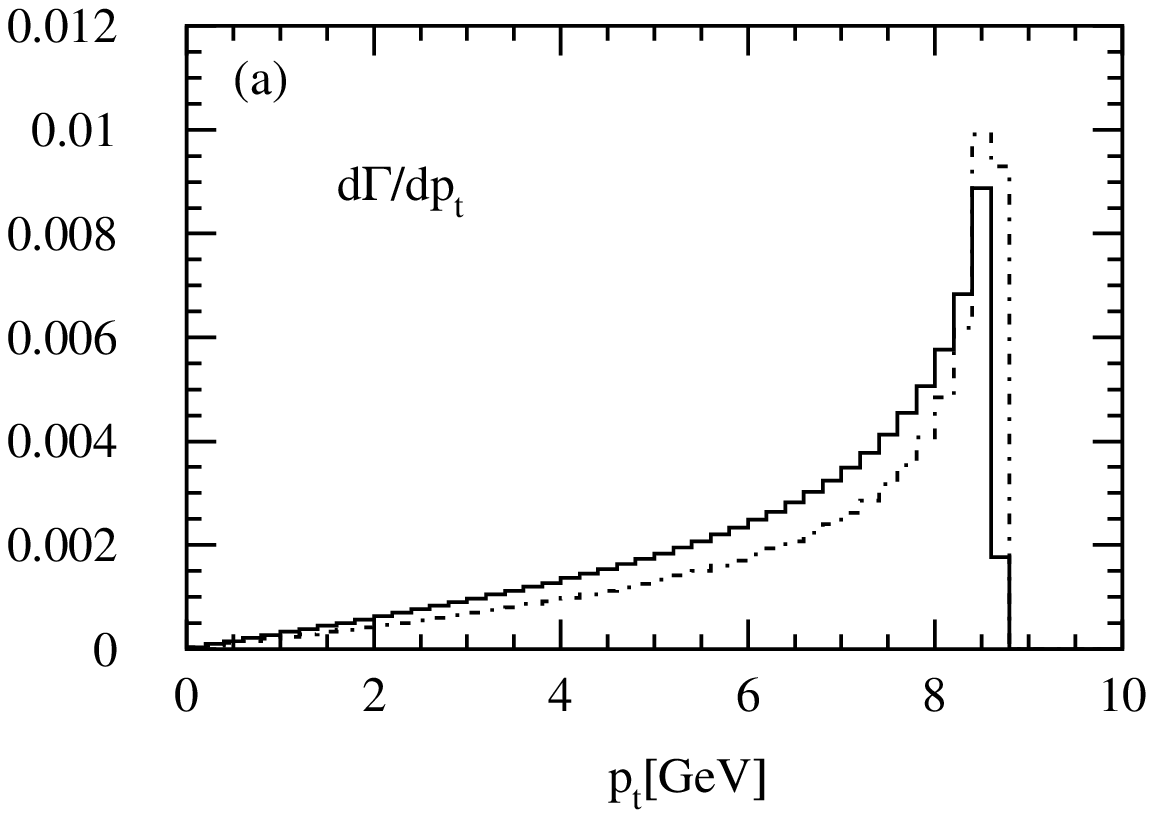,%
bbllx=50pt,bblly=120pt,bburx=285pt,bbury=460pt,angle=270,width=8.25cm}
\epsfig{file=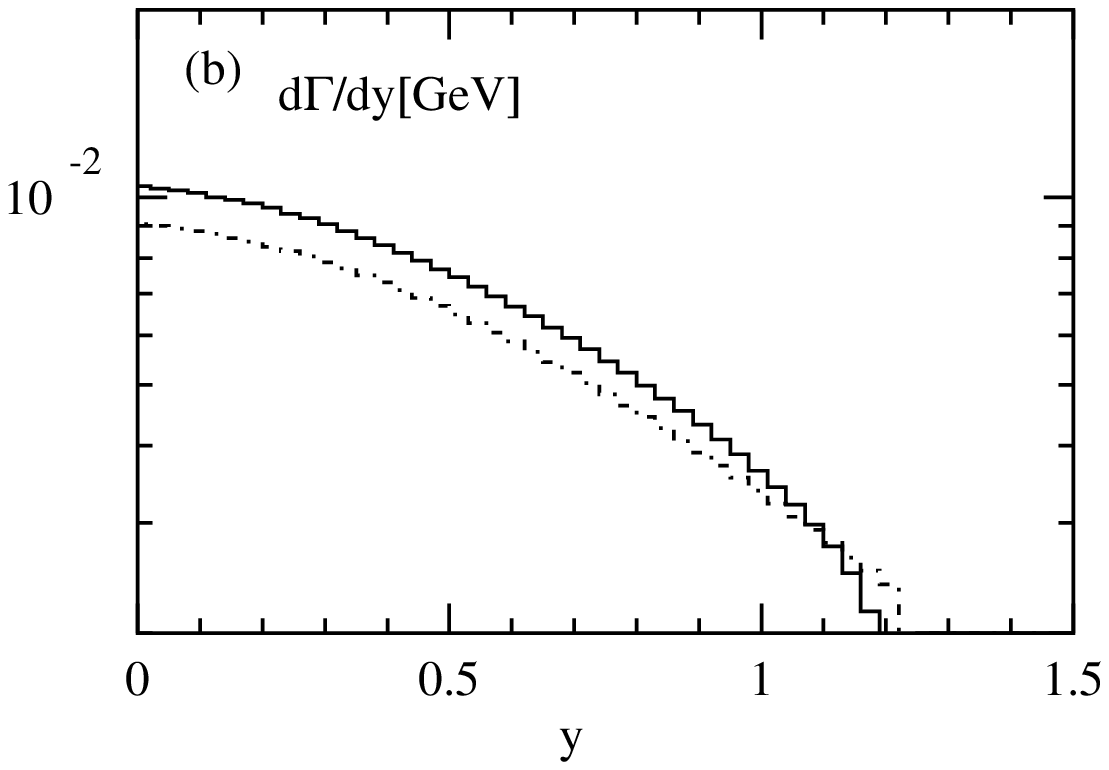,%
bbllx=50pt,bblly=120pt,bburx=285pt,bbury=460pt,angle=270,width=8.25cm}
\caption[dum]{\label{pperp-fig1}{\small{
Differential decay widths at Born (dotted-dashed)
and NLO (solid) levels, with parameters $s=400$ GeV$^2$,  
$m=5$ GeV, and $s_{\rm min}=0.001$ GeV$^2$ for differential variables
(a) transverse momentum $d\Gamma/dp_T$,
(b) rapidity $d\Gamma/dy$ [GeV].
}}}
\end{center}
\end{figure}

\begin{figure}[hbtp]
\begin{center}
\epsfig{file=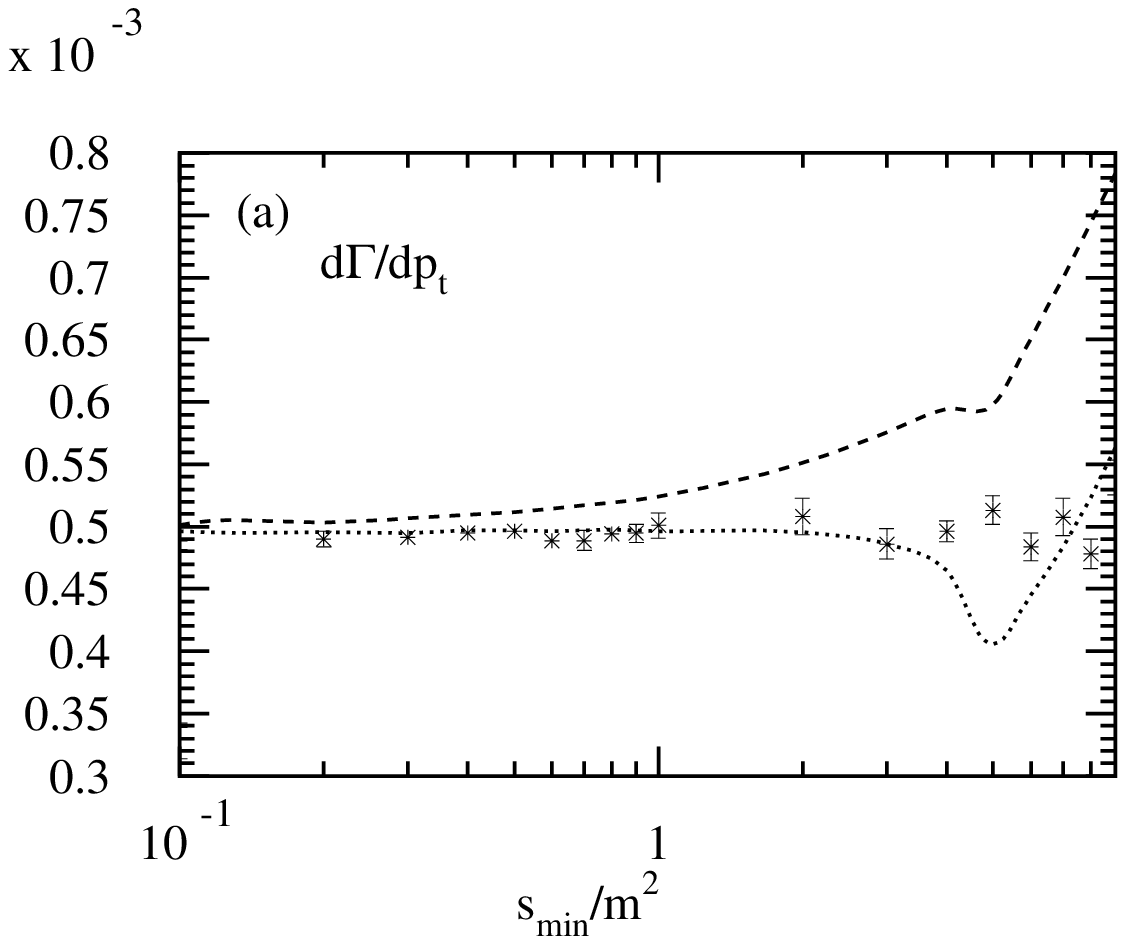,%
bbllx=50pt,bblly=120pt,bburx=285pt,bbury=460pt,angle=270,width=8.25cm}
\epsfig{file=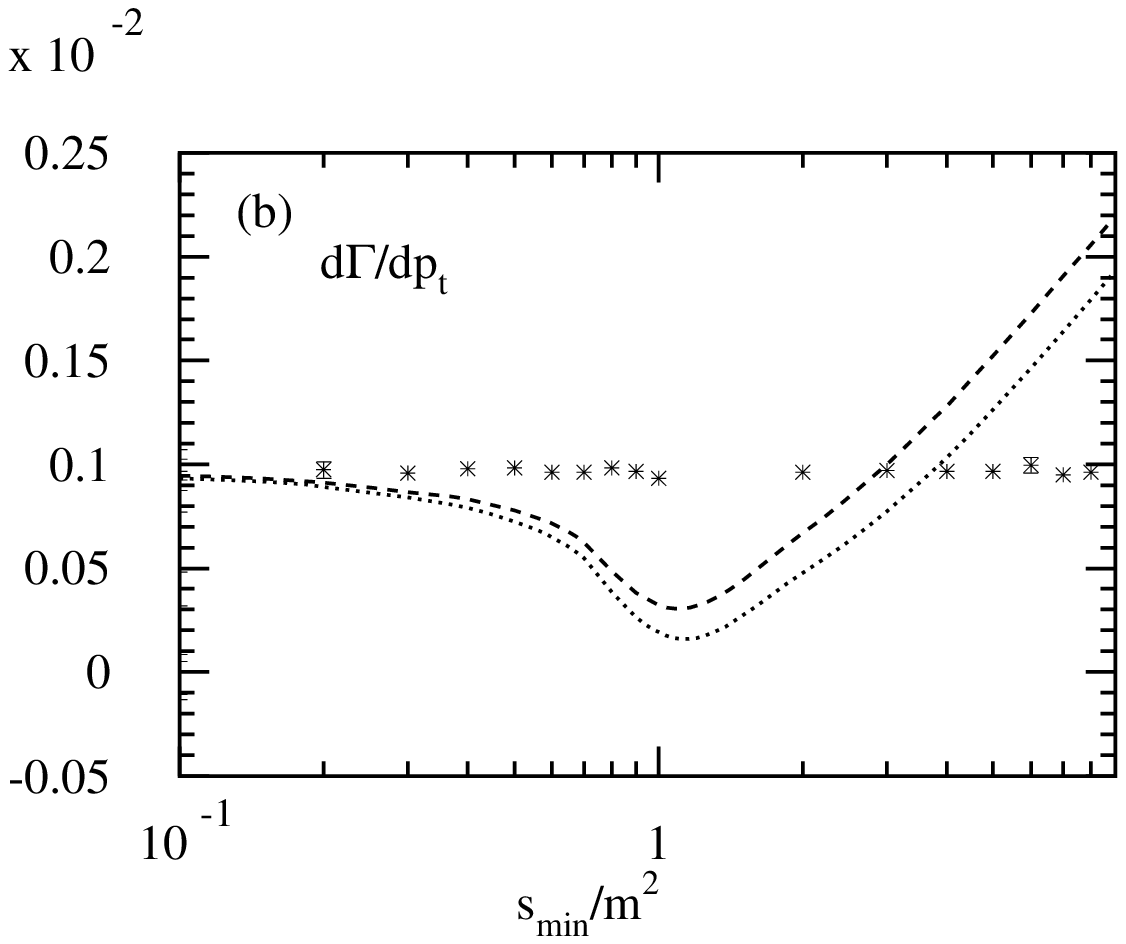,%
bbllx=50pt,bblly=120pt,bburx=285pt,bbury=460pt,angle=270,width=8.25cm}
\caption[dum]{\label{pperp-fig2}{\small{
      a) The $s_{\rm min}$-dependence of the one-loop contributions to
$d\Gamma/p_T$
for $p_T = 5$ GeV. We plot the results including 
the $T_1$ (dotted),
$T_1+T_2$ (dashed) and $T_1+T_2+T_3$ (individual points with error bars) terms.
      b) The $s_{\rm min}$-dependence of the one-loop contributions
to $d\Gamma/p_T$
for $p_T= 8$ GeV. Labels as in (a).
}}}
\end{center}
\end{figure}

\begin{figure}[hbtp]
\begin{center}
\epsfig{file=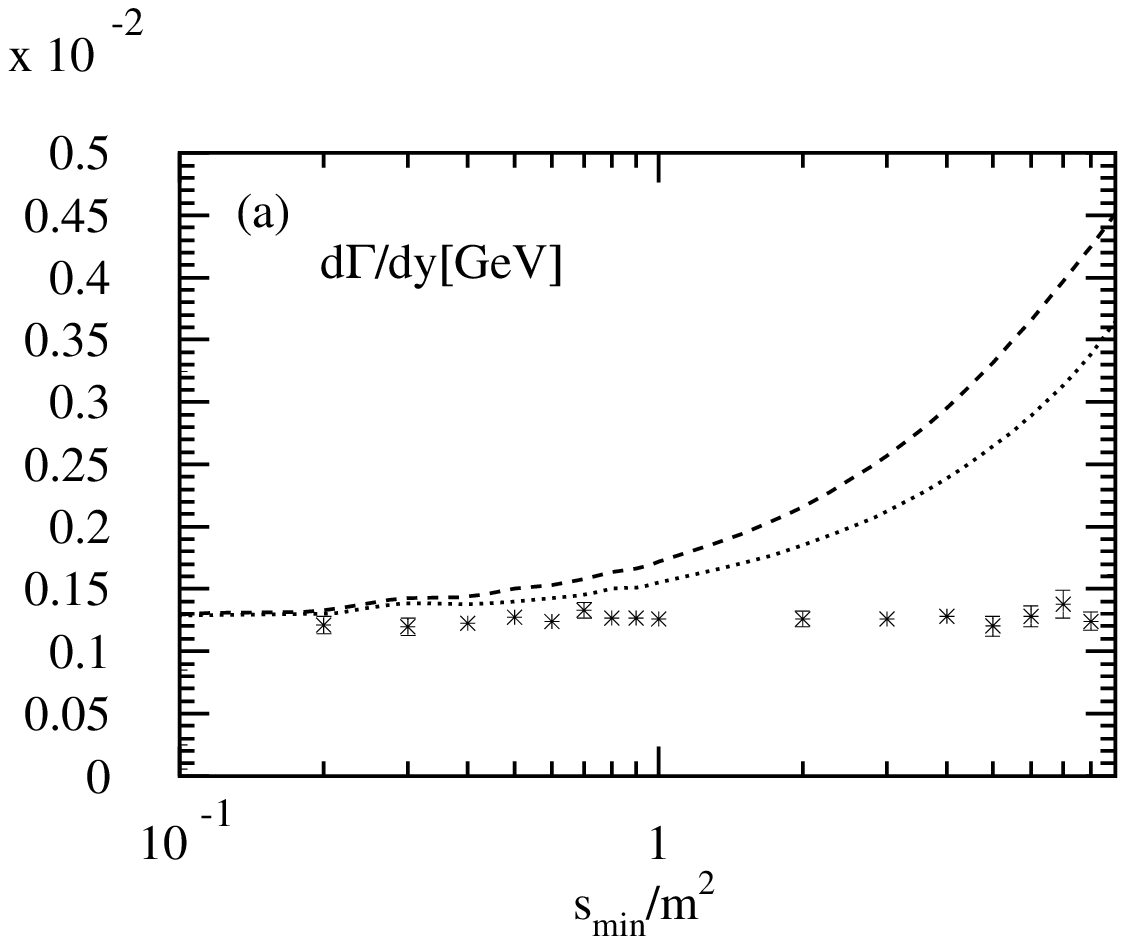,%
bbllx=50pt,bblly=120pt,bburx=285pt,bbury=460pt,angle=270,width=8.25cm}
\epsfig{file=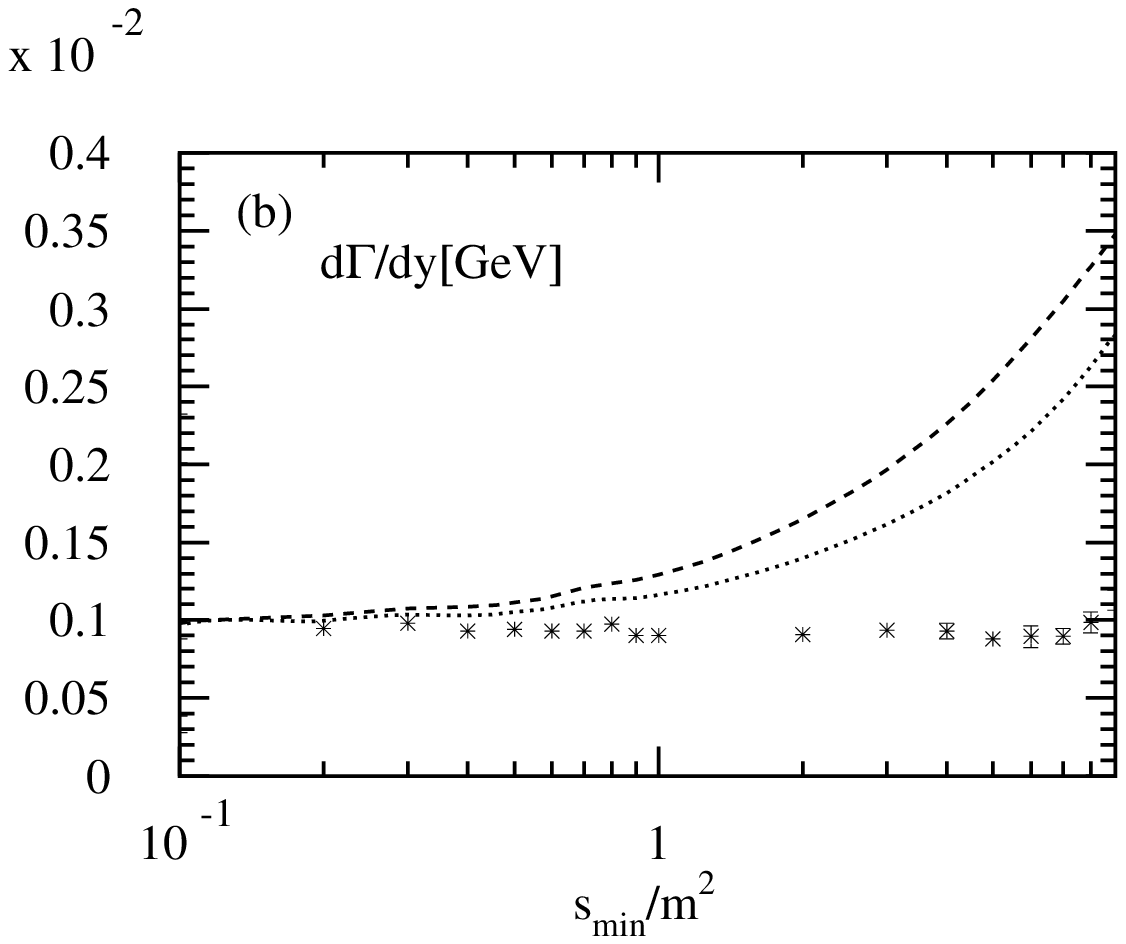,%
bbllx=50pt,bblly=120pt,bburx=285pt,bbury=460pt,angle=270,width=8.25cm}
\caption[dum]{\label{rap-fig2}{\small{
      a) The $s_{\rm min}$-dependence of of the one-loop contributions
to $d\Gamma/dy$
at $y=0.3$. Labels as in Fig.~\ref{pperp-fig2}.
      b) The $s_{\rm min}$-dependence of the one-loop 
contributions to $d\Gamma/dy$
at $y=0.6$. Labels as in Fig.~\ref{pperp-fig2}.
}}}
\end{center}
\end{figure}

Turning to distributions, we show in 
Fig.~\ref{pperp-fig1} the single heavy quark 
transverse momentum and rapidity distributions at 
a small value of $s_{\rm min} = 0.001{\rm GeV}^2$, 
computed with $T_1$ only. We see
the usual Jacobian peak near the kinematic maximum of the $p_T$ spectrum.
In Fig.~\ref{pperp-fig2} we plot the $s_{\rm min}$ dependence
of the one-loop contributions to $d\Gamma/p_T$ at two fixed 
values of $p_T$, one halfway and the other 
close to the kinematic maximum. The dip in the curves
is an artifact which arises because at that $s_{\rm min}$ and 
for the $p_T$ given, 
it is no longer kinematically possible for
the full phase space in Fig.~\ref{dahlitz-fig1} to contribute.
Note that the dip disappears for the exact $T_1+T_2+T_3$ case.
Similar results are shown for the 
heavy quark rapidity distributions in Fig.~\ref{rap-fig2}
(where we show only the positive-rapidity part of the distribution).
These figures show that the freedom to choose $s_{\rm min}$
when including all $T_i$ persists for distributions. 
\section{Dipole Subtraction}
\label{sec:dipole-subtraction}

In the dipole formalism one subtracts a suitable term from the real emission
part and adds it again to the virtual correction after having performed
one phase-space integration. 
The subtraction term consists of a sum of dipoles, each of which can be viewed
as an emitter-spectator-antenna radiating a third particle.
In the case at hand, there are only two dipoles. In one of these
the heavy quark constitutes
the emitter with the antiquark being the spectator.
The second dipole has the roles of the quark and antiquark exchanged.
The matrix element to be subtracted from the real emission part reads:
\begin{eqnarray}
  \label{eq:14}
\lefteqn{
\left|{\cal M}_{\rm A}\right|^2 = 2 C_F g_s^2 \, \left|{\cal M}_{\rm Born}\right|^2
 \frac{1}{r_0 r \sqrt{(1-r)(1-r_0 r)}} 
}\\
 & & \times 
 \left\{
  \frac{1}{s_{13}} \left[ 2(1-r_0r) -(1-r_0) - \frac{1-r_0}{1-u_0u} \right]
  + \frac{1}{s_{23}} \left[ 2(1-r_0r) -(1-r_0) - \frac{1-r_0}{1+u_0u} \right]
  \right\}\nonumber
\end{eqnarray}
Here
\begin{eqnarray}
  \label{eq:18}
r_0 = \beta^2, \; r = \frac{s_{13}+s_{23}}{s-4m^2},\; 
u_0 = \sqrt{\frac{r_0(1-r)}{1-r_0r}}, \;
u = - \frac{1}{u_0} \frac{s_{13}-s_{23}}{s_{13}+s_{23}}.
\end{eqnarray}
This contribution is then integrated over the dipole phase space and added
to the virtual corrections. The integrated version reads:
\begin{eqnarray}
  \label{eq:19}
\lefteqn{
\int d\,\mbox{PS}_{\rm dipole}
\left|{\cal M}_{\rm A}\right|^2 } \nonumber \\
 & = & C_F \frac{g_s^2}{4\pi^2} \frac{1}{\Gamma(1-\eps)} 
   \left( \frac{4\pi \mu^2}{s} \right)^\eps  \, \left|{\cal M}_{\rm Born}\right|^2
 \left\{
  \frac{1}{\eps} \left( 1 - \frac{1}{2} \frac{1+r_0}{\sqrt{r_0}} 
                            \ln \frac{1+\sqrt{r_0}}{1-\sqrt{r_0}} \right) 
 \right. \nonumber \\
 & &
  - 2 \ln r_0 
 - \ln^2\left(\frac{1+\sqrt{r_0}}{1-\sqrt{r_0}}\right) +\frac{1}{\sqrt{r_0}}\ln\left(\frac{1+\sqrt{r_0}}{1-\sqrt{r_0}}\right)
 \nonumber \\
 & & \left.
 - \frac{1+r_0}{2\sqrt{r_0}} 
   \left( \mbox{Li}_2\left(\sqrt{r_0}\right) - \mbox{Li}_2\left(-\sqrt{r_0}\right)
          +2\;\mbox{Li}_2\left(\frac{1+\sqrt{r_0}}{2}\right)-2\;\mbox{Li}_2\left(\frac{1-\sqrt{r_0}}{2}\right) 
 \right. \right. \nonumber \\
 & & \left. \left.
          +\mbox{Li}_2\left(\frac{\sqrt{r_0}-1}{2\sqrt{r_0}}\right)
          -\mbox{Li}_2\left(\frac{\sqrt{r_0}-1}{\sqrt{r_0}}\right)
          +\mbox{Li}_2\left(\frac{1}{1+\sqrt{r_0}}\right)
          -\mbox{Li}_2\left(\frac{1-\sqrt{r_0}}{1+\sqrt{r_0}}\right)
 \right. \right. \nonumber \\
 & & \left. \left.
          -2\ln r_0 \ln\left(\frac{1+\sqrt{r_0}}{1-\sqrt{r_0}}\right)
          +\ln2 \ln \frac{\sqrt{r_0}}{1+\sqrt{r_0}}
          +\frac{1}{2} \ln^2 2
 \right. \right. \nonumber \\
 & & \left. \left.
          + \ln(1-\sqrt{r_0}) \ln \left( \frac{1+\sqrt{r_0}}{\sqrt{r_0}} \right)
          + \frac{1}{2} \ln^2(1+\sqrt{r_0}) 
          - \frac{1}{2} \ln^2(1-\sqrt{r_0})
   \right) 
 \right\} 
 \nonumber \\ & & + O(\eps)
\end{eqnarray}
The poles in $\eps$ cancel against those 
of the virtual corrections.

We do not show separate results for the dipole method,
which is exact and independent of any theoretical cut-off
parameter. Numerical results for the dipole method in
comparison to the PSS method can be found in the next section.

\section{Comparisons of PSS and dipole subtraction}
\label{sec:comp-pss-dipole}

In this section we perform some numerical comparisons between
the two methods for the process at hand. We use as 
phase space measure the expression (\ref{eq:3}).
The integrations over its variables are performed 
using the well-known Monte Carlo iterative integration routine 
VEGAS \cite{Lepage:1978sw}. We note that we found similar 
results when we used (\ref{eq:8}), generating
the 4-vectors via a cascade algorithm. 
This required using more random number points in
order to achieve the same accuracy.

The PSS method is relatively easy to implement, 
with little analytical calculation,
at the expense of requiring cancellations between
large numbers (for small $s_{\mathrm{min}}$) or
having multiple negative contributions (for large $s_{\mathrm{min}}$
when including $T_1$, $T_2$ and $T_3$).
Since the dipole method requires more analytical preparation
work to be implemented, we expect  it to show
better numerical integration in the Monte
Carlo program. We will see that 
this expectation is borne out by our results.

Our first comparison addresses the relative accuracy achieved
in the computation of the inclusive cross section 
as a function of the number of points, for 20
iterations, of which we use the first five to set the
VEGAS grid \cite{Lepage:1978sw}, leaving a sample of $N=15$ results. 
For each method, we perform separate runs for the 
$O(\alpha_s)$ 2-particle and $O(\alpha_s)$ 3-particle 
contributions, and combine them for each iteration,
leading to 15 results $r_i$. 
The mean result $r$ and its error $\delta r$ are then computed as 
\begin{equation}
  \label{eq:7}
 r = \frac{1}{N}\sum_{i=1}^{N} r_i,\qquad 
 \delta r =  \sqrt{\frac{1}{N}\frac{\sum_{i=1}^N \left(r_i-r\right)^2}{N-1}}\,.
\end{equation}
The results for this comparison are given in Table \ref{tab:1}.
We note that the PSS method suffers further penalties
in accuracy and efficiency if the value of $s_{\mathrm{min}}$ is 
chosen so large that the $T_2$ and $T_3$ become necessary;
in particular the $T_3$ contribution requires generating the soft
phase space measure, and 
involves the difference of two phase space measures which 
are very similar in magnitude for small 
values of the soft invariants, cf. Eq.~(\ref{eq:T3}).
\begin{table}[tbp]
\begin{center}
\begin{tabular}{|c|c|c|} \hline
\multicolumn{3}{|c|}{$s=400$ GeV$^2$} \\ \hline
points & DIP & PSS   \\ \hline
1000 & 0.04\%  & 1\% \\ \hline
10000 & 0.009\%  & 0.3\% \\ \hline
100000 & 0.003\%  & 0.1\% \\ \hline
\end{tabular}
\caption[]{{\small{Accuracy $\delta r/r$ of the inclusive cross
      section attained for a given number
      of points per iteration in
the two methods. The same phase space and random number generators 
are employed. The PSS results use the $T_1$ contribution only,
with $s_{\rm min} = 0.001 {\rm GeV}^2$.}}}
\label{tab:1}
\end{center}
\end{table}   

Our second comparison addresses the efficiency
in the computation of the inclusive cross section 
as a function of the number of iterations, for
$10^4$ random number points.
\begin{table}[tbp]
\begin{center}
\begin{tabular}{|c|c|c|} \hline
iteration & DIP & PSS   \\ \hline
1 & 0.1\%  & 100\% \\ \hline
2 & 0.09\%  & 70\% \\ \hline
3 & 0.06\%  & 10\% \\ \hline
\end{tabular}
\caption[]{{\small{Comparison of the two methods as to the
approximate relative deviations of their first three (grid-setting)
iterations from the final mean 
(computed starting from the fifth iteration), for the case of the
inclusive cross section.
The same phase space and random number generators 
are employed, at $s=400$ GeV$^2$. The PSS results use the $T_1$ contribution only,
with $s_{\rm min} = 0.001 {\rm GeV}^2$.}}}
\label{tab:2}
\end{center}
\end{table}   
We see that the dipole method reaches a given accuracy with less
iterations. 

Next we compare the efficiency of these methods to 
compute transverse momentum and rapidity distributions.
\begin{figure}[hbtp]
\begin{center}
\epsfig{file=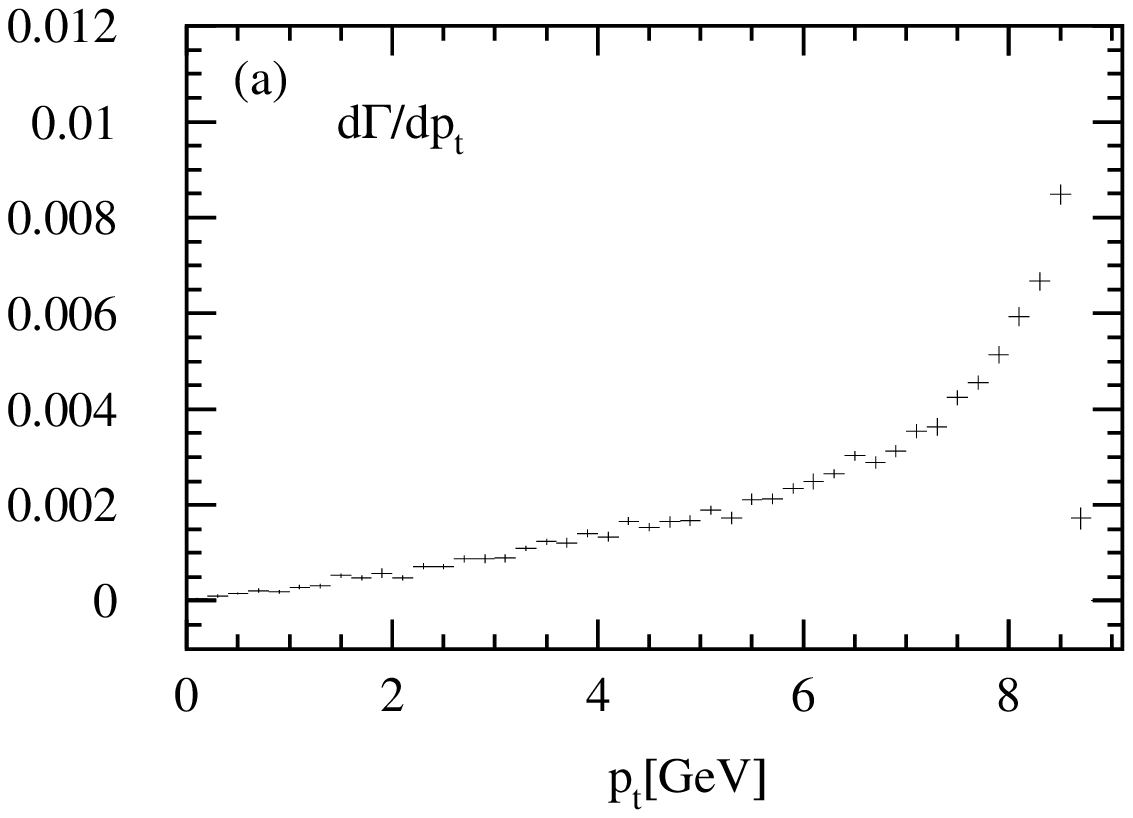,%
bbllx=50pt,bblly=120pt,bburx=285pt,bbury=460pt,angle=270,width=8.25cm}
\epsfig{file=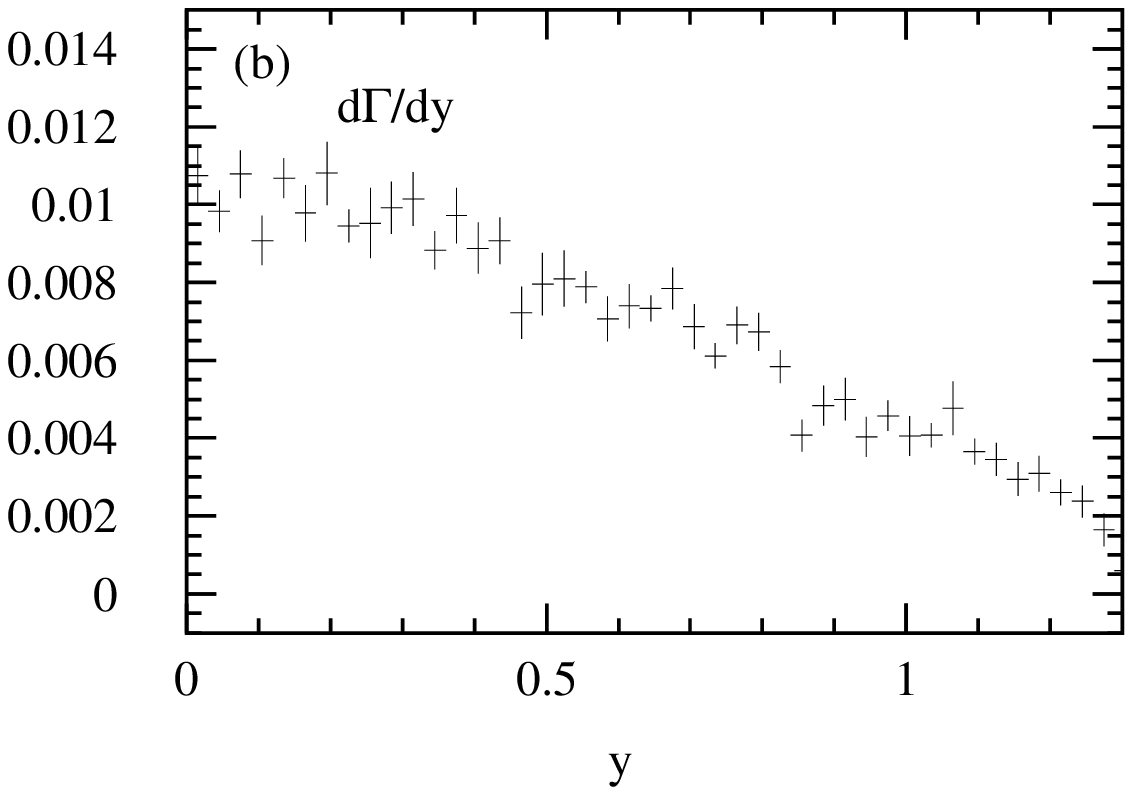,%
bbllx=50pt,bblly=120pt,bburx=285pt,bbury=460pt,angle=270,width=8.25cm}
\caption[dum]{\label{discomp-fig1}{\small{
Differential decay widths at NLO level, with parameters $s=400$ GeV$^2$,  
$m=5$ GeV, for the phase space slicing method
(at $s_{\rm min}=0.001 {\rm GeV}^2$) for differential variables
(a) transverse momentum $d\Gamma/dp_T$ 
(b) rapidity $d\Gamma/dy$ [GeV]}}}
\end{center}
\end{figure}
\begin{figure}[hbtp]
\begin{center}
\epsfig{file=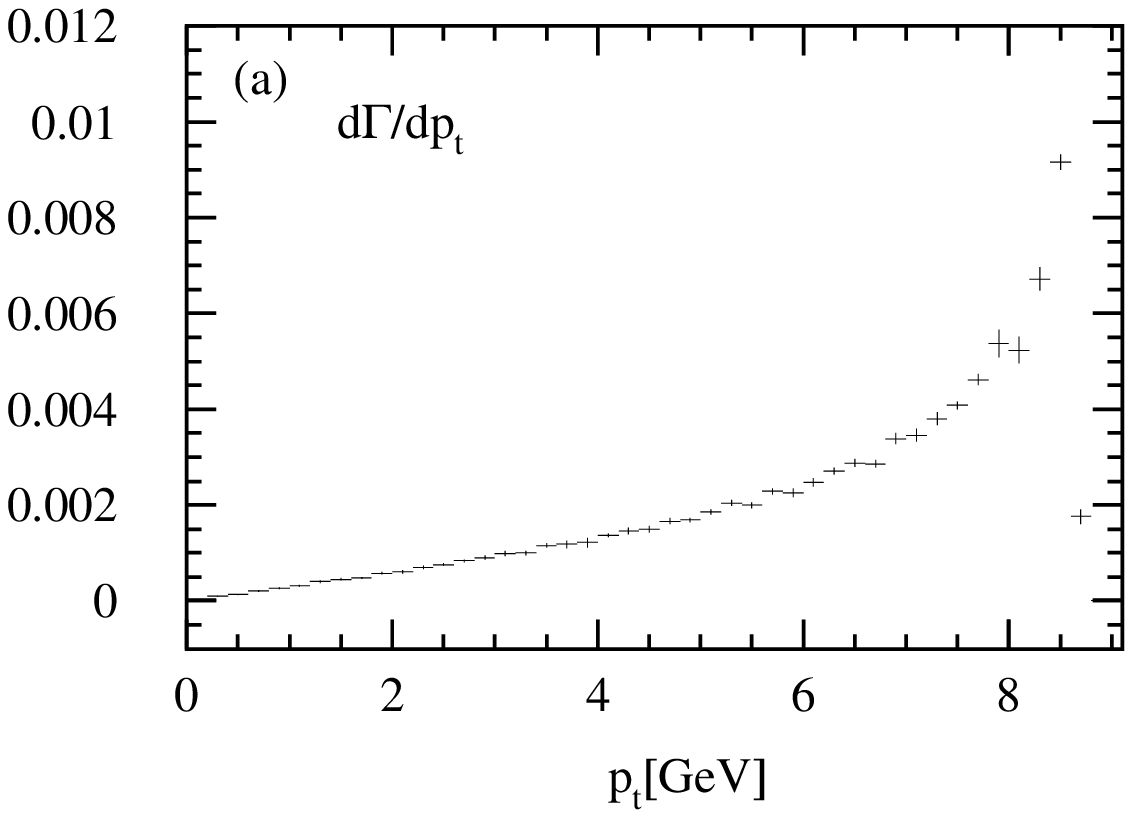,%
bbllx=50pt,bblly=120pt,bburx=285pt,bbury=460pt,angle=270,width=8.25cm}
\epsfig{file=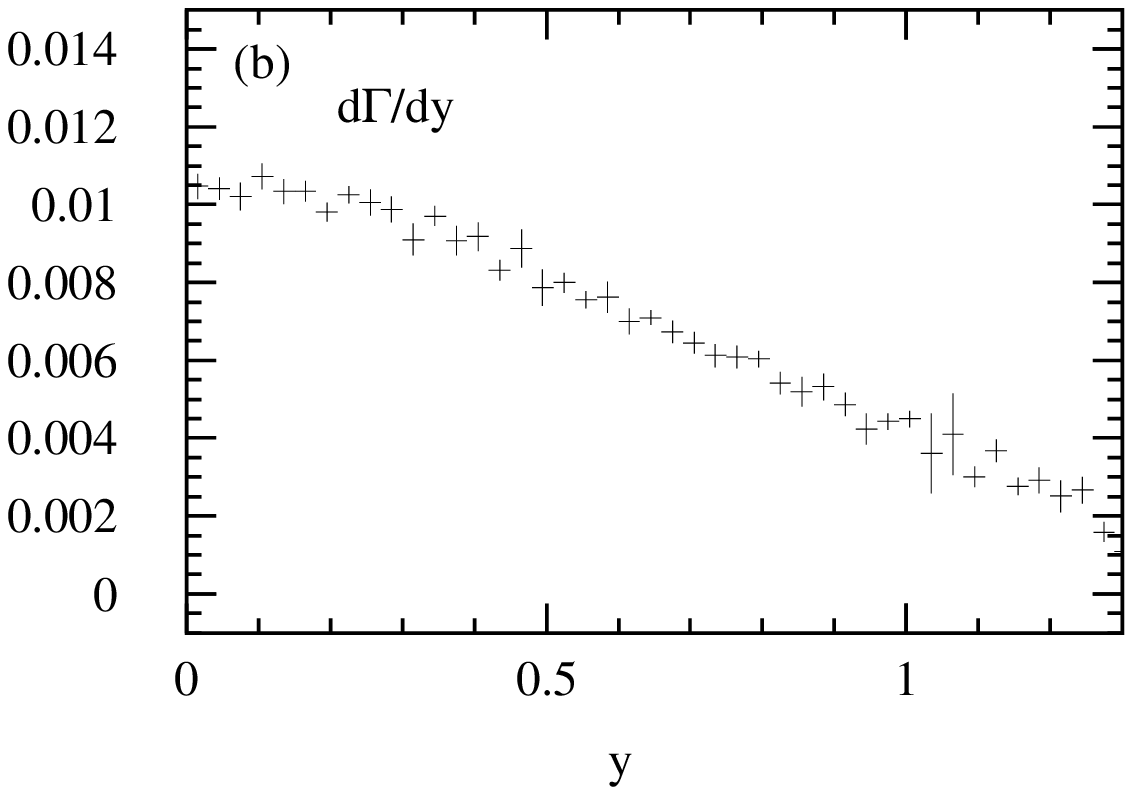,%
bbllx=50pt,bblly=120pt,bburx=285pt,bbury=460pt,angle=270,width=8.25cm}
\caption[dum]{\label{discomp-fig2}{\small{
Differential decay widths at NLO level, with parameters $s=400$ GeV$^2$,  
$m=5$ GeV, for the dipole method for differential variables
(a) transverse momentum $d\Gamma/dp_T$ 
(b) rapidity $d\Gamma/dy$ [GeV] }}}
\end{center}
\end{figure}
As before, the first 5 of 20 iterations are used solely for
grid-setting, with $10^4$ points per iteration.
The values and their errors for each bin are computed according to (\ref{eq:7}).
We see that the dipole methods produces somewhat smaller errors, with
slightly less bin to bin variations. When we increased the
number of points, we saw that both methods perform not too differently.
We also noticed that this loss of superiority is progressive with
the number of bins. This suggests that parts of the positive
and negative contributions end up in different bins.
To test this idea, we performed a simple smearing where each
event with weight $w$ that would normally end up in bin $i$ is distributed
in bins $i-1,i,i+1$, each with weight $w/3$. We found that this
smearing indeed reduced the errors somewhat, but in about equal measure for
both methods.

Finally, we compared the accuracies of the methods in the large $s$
limit for the inclusive cross section, where the heavy quarks
become effectively massless.  For the PSS method we
investigated how to choose $s_{\mathrm{min}}$ in order to minimize
the calculation error. We found that $s_{\mathrm{min}}$ is best
chosen not too small (which would lead to large numerical
cancellations), and as a fraction of $s$ between 0.01 and 0.1.
For large $s_{\mathrm{min}}$ this may require the inclusion of the $T_2$ and $T_3$
terms, which contribute about 10\% to the cross section for 
$s_{\mathrm{min}} = 0.1\,s$ at $s=250000$ GeV$^2$. For $s=250000$ GeV$^2$
and $s_{\mathrm{min}} = 0.01\,s$ their contribution is only 2\%,
with a slightly larger total error for the same number of points.
We found similar results keeping $s$ fixed and letting $m$
become smaller. For the dipole method we found that it has
consistently better accuracy than the PSS method in these limits.
In general in the high-energy limit, both these methods lead to
cancellations between contributions with $+\ln(s/m^2)$
and $-\ln(s/m^2)$ terms, which is not advantageous 
numerically. Therefore a method which avoids such logarithms
could be desirable \cite{othermassdipole}.

\section{Conclusions}
\label{sec:conclusions}

In this paper we have compared the
accuracy and efficiency of two general-purpose methods to
compute NLO heavy quark production cross sections,
for a very simple case. We found the dipole method \cite{Catani:1997vz,Phaf:2001gc}, 
while involving additional analytical work, to be superior in 
efficiency and accuracy. 
A similar conclusion was reached by 
Dittmaier \cite{Dittmaier:1999mb} who compared 
his method with a slicing calculation
for a number of electroweak cross sections.

The phase space slicing method
\cite{Giele:1992vf,Giele:1993dj,Keller:1998tf}, which is easy to 
use and minimizes analytical work, can be extended
\cite{Kilgore:1997sq} to become fully independent of the slicing
parameter, which we demonstrated in this paper for
the reaction at hand. 
Although our case-study involves only the simplest of
heavy quark production processes, it is, we believe
a useful first step toward gaining numerical experience with
general methods for constructing NLO Monte Carlo programs
for heavy quark production. 
Moreover, such experience gained at NLO is likely to be 
very valuable when these methods are generalized for NNLO
cross sections.

\subsection*{Acknowledgements}
\label{sec:acknowledgements}

The work of T.O.E., E.L. and L.P. is supported by the Foundation for 
Fundamental Research of Matter (FOM) and the National Organization 
for Scientific Research (NWO).

\def\appendix{\setcounter{section}{0} \setcounter{equation}{0}
  \def\thesection{\Alph{section}}
  \def\theequation{\Alph{section}.\arabic{equation}}}
\appendix

\section{A subtlety in Phase Space Slicing}

In this appendix we mention a subtlety in implementing PSS, which is well known to
experts in the field, but not readily found in the literature. 
It does not come into play for our simple case study, but
does for amplitudes in which more than one color structure is present.
The correct implementation of phase space slicing requires that the
real emission amplitude be decomposed 
into pieces with a unique singularity structure and that the slicing procedure
be defined for each piece separately.

We discuss the point for a simplified example, where we take al
particles to be massless. We assume that
the real emission amplitude is given by
\bq
\left| {\cal M}_4 \right|^2 & = & \frac{1}{s_{13} s_{34} s_{24}} 
                                + \frac{1}{s_{14} s_{34} s_{23}}.
\eq
This simple example has the singularity structure for the leading-color part of the
$\gamma^\ast \rightarrow q\bar{q} g g$ amplitude.
The correct way to implement phase space slicing treats each singularity structure
separately and the resolved contribution from the real emission amplitude reads therefore
\bq
& &
\int d\,\mbox{PS}_{4} \frac{1}{s_{13} s_{34} s_{24}}
 \theta\left( s_{13} - s_{\rm min} \right) 
 \theta\left( s_{34} - s_{\rm min} \right) 
 \theta\left( s_{24} - s_{\rm min} \right) 
\nonumber \\
& &
 +
\int d\,\mbox{PS}_{4} \frac{1}{s_{14} s_{34} s_{23}}
 \theta\left( s_{14} - s_{\rm min} \right) 
 \theta\left( s_{34} - s_{\rm min} \right) 
 \theta\left( s_{23} - s_{\rm min} \right).
\eq
The incorrect method cuts out all possible singularities from the amplitude and uses
the expression 
\bq
& & 
\int d\,\mbox{PS}_{4} \left(
  \frac{1}{s_{13} s_{34} s_{24}}
+ \frac{1}{s_{14} s_{34} s_{23}}
                      \right)
 \theta\left( s_{13} - s_{\rm min} \right) 
 \theta\left( s_{34} - s_{\rm min} \right) 
 \theta\left( s_{24} - s_{\rm min} \right) 
\nonumber \\
& & \hspace{6cm} \cdot \;
 \theta\left( s_{14} - s_{\rm min} \right) 
 \theta\left( s_{23} - s_{\rm min} \right) 
\eq
for the resolved contribution. The difference between the correct implementation and
the incorrect one consists of terms of the form
\bq
\int d\,\mbox{PS}_{4} 
  \frac{1}{s_{13} s_{34} s_{24}}
 \theta\left( s_{13} - s_{\rm min} \right) 
 \theta\left( s_{34} - s_{\rm min} \right) 
 \theta\left( s_{24} - s_{\rm min} \right) 
 \theta\left( s_{\rm min} - s_{14} \right) 
 \theta\left( s_{23} - s_{14} \right) 
\eq
together with three similar terms, obtained by exchanging $3 \leftrightarrow 4$ in
the matrix element and the slicing procedure.
Contrary to naive expectations, these terms do not vanish in the limit
$ s_{\rm min} \rightarrow 0$, but give a constant contribution.


\end{document}